\documentclass{ws-ijmpa}

\begin{document}

\markboth{J. W. Moffat}{GRAVITATIONAL INSTABILITY OF THE VACUUM
AND THE COSMOLOGICAL CONSTANT PROBLEM}

\title{GRAVITATIONAL INSTABILITY OF THE VACUUM AND THE COSMOLOGICAL CONSTANT
PROBLEM}

\author{J. W. MOFFAT}

\address{{\it Perimeter Institute for Theoretical Physics, Address\\31 Caroline St. North, Waterloo,
Ontario, Canada}}

\address{{\it Department of Physics, University of Waterloo, Address\\Waterloo,
Ontario, Canada}}

\maketitle

\begin{abstract}
A mechanism for suppressing the cosmological constant is
described, using a superconducting analogy in which fermions
coupled perturbatively to gravitons are in an unstable false
vacuum. The coupling of the fermions to gravitons and a screened
attractive interaction among pairs of fermions generates fermion
condensates with zero momentum and a phase transition induces a
non-perturbative transition to a true vacuum state. This produces
a positive energy gap $\Delta$ in the vacuum energy identified
with $\sqrt{\Lambda}$, where $\Lambda$ is the cosmological
constant. In the strong coupling limit, a large cosmological
constant induces a period of inflation in the early universe,
followed by a weak coupling limit in which $\sqrt{\Lambda}$
vanishes exponentially fast as the universe expands due to the
dependence of the energy gap on the density of Fermi surface
fermions, predicting a small cosmological constant in the early
universe.
\end{abstract}

\section{Cosmological Constant Problem, Vacuum Energy and
Inflation}

It is generally agreed that the cosmological constant problem
(CCP) is one of the most severe problems facing modern particle
and gravitational physics. It is believed that its solution could
significantly alter our understanding of particle physics and
cosmology. The problem has become more severe with the discovery
that the universe is accelerating and that the so-called ``dark
energy'' can be explained by a non-zero and small cosmological
constant.

The cosmological constant problem centers around two questions:
why is the cosmological constant observed to be so small and why
does it dominate the universe today~\cite{Weinberg,Straumann}.
Many investigators have proposed solutions ranging from a rolling
scalar field to the anthropic principle. However, it can be argued
that the greatest irony of the cosmological constant problem is
the inflationary paradigm. The onset of the inflationary epoch is
completely dominated with a fluid that is either a pure
cosmological constant or a scalar field whose potential is
extremely flat, mimicking a cosmological constant. Clearly this
period of inflation has to end for successful structure formation.
However, whatever mechanism is responsible for this exit from
inflation should be a clue as to how the cosmological constant is
relaxed at all times including today.

We shall propose a mechanism to solve the cosmological constant
problem, based on a simple idea analogous to the microscopic
realization of superconductivity~\cite{Moffat}. We argue that the
perturbative vacuum state in the presence of a cosmological
constant is gravitationally unstable and it is energetically
favorable for the vacuum associated with the effective
cosmological constant to release all of its energy into the
production of fermion condensates. The formation of condensates
leads to a non-perturbative true ground state. The
non-perturbative gravitational instability of the vacuum due to a
transition of fermions to a superfluid of condensates in the early
universe, produces a large cosmological constant which induces a
de Sitter expansion and inflation, followed by an exponential
suppression of the cosmological constant as the universe expands.

We can define an effective cosmological constant
\begin{equation}
\lambda_{\rm eff}=\lambda_0+\lambda_{\rm vac},
\end{equation}
where $\lambda_0$ is the ``bare'' cosmological
constant in Einstein's classical field equations,
and $\lambda_{\rm vac}$ is the contribution that arises from the
vacuum density $\lambda_{\rm vac}=8\pi G\rho_{\rm vac}$.

Already at the standard model electroweak scale $\sim 10^2$ GeV, a
calculation of the vacuum density $\rho_{\rm vac}$, based on local
{\it perturbative} quantum field theory results in a discrepancy
of order $10^{55}$ with the observational bound
\begin{equation}
\rho_{\rm vac} \leq 10^{-47}\, ({\rm GeV})^4 \sim (10^{-3}\,eV)^4.
\end{equation}
If the vacuum energy is the dark energy, then $\rho_{\rm vac}\sim
(10^{-3}\, eV)^4$. {\bf There is an egregious discrepancy between
the particle physics estimate of $\rho_{\rm vac}$ and the
cosmological observation}. We are faced with a severe fine-tuning
problem of order $10^{55}$, since the virtual quantum fluctuations
giving rise to $\lambda_{\rm vac}$ must cancel $\lambda_0$ to an
unbelievable degree of accuracy. This is the ``particle physics''
source of the cosmological constant problem.

In order to achieve at least 60 e-folds of inflation, we need to
postulate a large vacuum energy initially in the early universe,
which generates a de Sitter phase of inflation
\begin{equation}
a(t)\sim \exp(H_0t)\sim \exp(\sqrt{\Lambda/3}t).
\end{equation}
A major problem for inflationary models is to explain why the
cosmological constant (vacuum energy) is very small in the present
universe. The usual statement is that the vacuum energy decays
into particles but this implies a severe fine tuning and it is not
accompanied by a satisfactory particle physics description of the
mechanism that underlies the suppression of the initial large
vacuum energy.

We will argue that the perturbative vacuum for fermions
interacting with gravitons is unstable to the production of
massive condensate states, transmuting the large vacuum energy
into rest mass.  In the case of inflation, these states are the
Goldstone bosons corresponding to the broken de Sitter space-time
symmetry which commences with a lower cosmological constant. Let
us recall that this situation is similar to the vacuum instability
phenomena of superconductivity in condensed matter systems and QCD
at finite density. In the BCS~\cite{Bardeen} theory of
superconductivity, the naive ground state associated with the
Fermi surface is unstable in the presence of an attractive
electron-phonon interaction, which acts to screen the repulsive
Coulomb interaction.  This instability is quantified by a gap
equation, which is an energy gap $\Delta$ between the
superconducting true ground state and the ordinary conduction
state.

\section{Vacuum Instability and Fermion Condensates}

We propose a {\it non-perturbative} mechanism to solve the
cosmological constant problem, based on an analogy with the
microphysical realization of superconductivity~\cite{Taylor}. We
argue that in the absence of gravitational interactions between
fermions, $\Lambda=0$ and when the fermions exchange gravitons
Minkowski spacetime becomes unstable. A {\it non-perturbative}
phase transition to a true vacuum state occurs when the
gravitational interaction is taken into account. The fermions form
condensates with zero momentum due to the weak gravitational
interaction and a screened long-range attractive interaction among
the pairs of fermions.

We shall describe the phase transition to fermion condensates
using a non-relativistic toy model. The Hamiltonian takes the BCS
form with ${\bf k}=-{\bf k}$:
\begin{equation}
H=\sum_{ks}{\cal E}_kc_{ks}^\dagger c_{ks}
-\frac{1}{2}\sum_{k,k's}V_{kk'}c_{k's}^\dagger c_{-k's}^\dagger
c_{-k's}c_{ks}.
\end{equation}

We perform the transformation to new operators
\begin{equation}
b_k=u_kc_k-v_kc^\dagger_{-k},\quad
b_{-k}=u_kc_{-k}+v_kc^\dagger_k,
\end{equation}
where the bs satisfy anti-commutation relations and
$u^2_k+v_k^2=1$. The fermion number operators are $n_k=b_k^\dagger
b_k$ and $n_{-k}=b_{-k}^\dagger b_{-k}$.

\section{Gap Equation and the Cosmological Constant}

We must now determine the ground state (vacuum) and set the
occupation numbers $n_k$ and $n_{-k}$ equal to zero. We need to
determine the energy gap $\Delta$ produced by the gap in the
vacuum energy in the phase transition to the fermion condensates.
The condensates are bound states due to the weak gravitational
interaction generated by the exchange of gravitons between
fermions and the screened attractive force. This will give
\begin{equation}
\Delta=\sqrt{\Lambda}.
\end{equation}

To find the minimum of the BCS energy, we diagonalize $H$ and
obtain the condition
\begin{equation} {\cal E}_k\biggl(\frac{1}{4}-x_k^2\biggr)^{1/2}+x_k
\sum_{k'}V_{kk'}\biggl(\frac{1}{4}-x_{k'}\biggr)^{1/2}=0,
\end{equation}
where $u_k=(\frac{1}{2}-x_k)^{1/2}$ and
$v_k=(\frac{1}{2}+x_k)^{1/2}$.

Let us define the quantity
\begin{equation}
\Delta_k=\sum_{k'}V_{kk'}\biggl(\frac{1}{4}-x_{k'}\biggr)^{1/2},
\end{equation}
giving
\begin{equation}
\Delta_k=\frac{1}{2}\sum_{k'}V_{kk'}\frac{\Delta_{k'}}{({\cal
E}^2_{k'} +\Delta_{k'}^2)^{1/2}}.
\end{equation}

We assume the simple model for the interaction matrix
\begin{equation}
V_{kk'}=V\quad {\rm if}\,\, \vert {\cal E}_k\vert < \omega_D,\quad
V_{kk'}=0\quad {\rm otherwise}.
\end{equation}

We obtain for the energy gap:
\begin{equation}
\Delta=\sqrt{\Lambda}=\frac{\omega_D}{\sinh[1/VD]},
\end{equation}
where $D$ is the fermion density of states defined by the Fermi
sphere for $N$ fermions by
\begin{equation}
n_f=\int_{k_0}^{k_f}d^3kD(k)
\end{equation}
and $\omega_D$ ($\hbar=1$) is the Debye energy.

The physical density of states behaves for an expanding universe
as
\begin{equation}
D_{\rm phys}(\omega_k)\sim a^3(t),
\end{equation}
so that as the universe inflates and $a(t)\rightarrow\infty$, we
have $D(\omega)\rightarrow \infty$. However, we assume as a first
approximation that $VD$ is constant and independent of $a(t)$.

We find that
\begin{equation}
V_{kk'q}=\frac{\vert M_q\vert^2\omega_D} {({\cal E}_k-{\cal
E}_{k-q})^2-\omega_D^2}\sim \frac{m_f}{M_f}
\biggl(\frac{NV_k}{\omega_D}\biggr)^2.
\end{equation}
Here, $m_f$ and $M_f$ denote the negatively charged condensate
fermion mass and the positively charged fermion ``ion'' mass,
respectively.

In the early universe there is an initial phase in which spacetime is flat
(Minkowski) and the fermions do not interact gravitationally. This phase is
unstable to gravitational interactions between fermions. There is a phase
transition to an inflating de Sitter vacuum, in which $VD\sim 1$
corresponding to a {\it strong coupling} limit and
\begin{equation}
\Delta_i=\sqrt{\Lambda}\sim \omega_D.
\end{equation}
In this phase
\begin{equation}
\Delta_i\sim \omega_D \sim M_{\rm PL},
\end{equation}
where $M_{\rm PL}$ is the Planck mass. As the
universe expands exponentially, a {\it weak coupling} limit
develops when
\begin{equation} VD\ll 1,
\end{equation}
which leads to an exponential suppression
\begin{equation}
\Delta_f=\sqrt{\Lambda}=2\omega_D\exp\biggl(-\frac{1}{VD}\biggr).
\end{equation}

We see that the fermion condensate phase can generate enough
inflation initially and then as the universe expands towards the
present epoch produce an exponential suppression of the
cosmological constant, leading to a zero or small value of
$\Lambda$ in the present universe.

The condensation energy for the weak coupling limit $VD\ll 1$ is
given by
\begin{equation}
{\cal E}_{\rm cond}\sim
-2\omega_D^2D\exp\biggl(-\frac{2}{VD}\biggr)\sim
-\frac{1}{2}D\Lambda^2.
\end{equation}
Thus, as the universe expands from its initial inflationary period
the number of fermions that is affected by the attractive
gravitational and screened interaction is a small fraction of the
total number of fermions in the universe. We note that
$\exp[-2/(VD)]$ has an essential singularity at $V=0$, which means
that while the function and its derivatives vanish as
$V\rightarrow +0$, they all become infinite as $V\rightarrow -0$.
This means that we cannot calculate ${\cal E}_{\rm cond}$ by using
perturbation theory.

What about the fluctuations that cause ultraviolet divergences? We
claim that these have to be subtracted out from the total vacuum
energy and do not play a fundamental role in the calculation of
the vacuum energy. This is the procedure used to calculate the
Casimir vacuum energy inside, for example, a spherical ball with
thin walls. The cosmological constant problem arises from physics
in the low-energy, infrared domain and the ultraviolet divergences
will be removed by a future self-consistent non-perturbative
quantum field theory and quantum gravity theory.

\section{Vacuum Instability in the de Sitter Phase}

Let us illustrate this mechanism with a toy model which
effectively models the formation of ``Cooper-pairs'' in a
relativistic context. Consider the following theory with an action
describing massless fermions coupled to gravity~\cite{Nambu}:
\begin{equation}
\label{NJL} {\cal L}= \sqrt{-g}[R +
\sum_{k=1}^{N}\bar{\psi}D_{a}\gamma^{a}\psi \;\; + \sum_{k=1}^{N}
\frac{G_0}{2N}(\bar{\psi}\psi\bar{\psi}\psi)],
\end{equation}
where $D_a$ is the covariant derivative with respect to the local
spin connection, $G_0$ is the ``bare'' gravitational coupling
constant, $N$ is the number of fermion species, and the third term
is a four-fermion interaction, which describes the relevant
graviton interaction between pairs of fermions on the Fermi
surface.

It is well-known that the physics described by (\ref{NJL}) is
equivalent to the following Lagrangian
\begin{equation}
{\cal L}= \sqrt{-g}[R + \bar{\psi}D_{a}\gamma^{a}\psi\;\; +
\bar{\psi}\phi\psi - \frac{N}{2G_0}\phi^{2}],
\end{equation}
where
$\phi=<\bar{\psi}\psi>$ is the condensate which forms from
fermion-graviton interactions.

As a simple first step, we will consider graviton exchange between
pairs of fermions by expanding about Minkowski spacetime:
\begin{equation}
g_{\mu\nu}=\eta_{\mu\nu}+h_{\mu\nu}+O(h^2),
\end{equation}
where $\eta_{\mu\nu}$ is the background Minkowski space metric:
$\eta_{\mu\nu}={\rm diag}(+1,-1,-1,-1)$. For a Dirac fermion, we
obtain for a momentum cutoff $K_c$ in the weak curvature limit
\begin{equation} i\gamma^\mu
p_\mu+\sqrt{\Lambda_0}+\Sigma(p,\sqrt{\Lambda},G_0,K_c)=0
\end{equation}
for $i\gamma^\mu p_\mu+\sqrt{\Lambda}=0$.

We have for a zero bare cosmological constant, $\Lambda_0=0$:
\begin{equation}
\sqrt{\Lambda}=\Sigma.
\end{equation}
For gravity for the
lowest-order loop we obtain
\begin{equation} \sqrt{\Lambda}
=G_0\sqrt{\Lambda}F(\sqrt{\Lambda},K_c),
\end{equation} where
$F(\sqrt{\Lambda},K_c)$ is the result of the momentum integration
of the Feynman fermion propagators and the cutoff is $K_c=M_{\rm
PL}$, where $M_{\rm PL}$ is the Planck mass. This has two
solutions: either $\sqrt{\Lambda}$ is zero or
\begin{equation}
\label{gap} \frac{1}{G_0}=F(\sqrt{\Lambda},K_c).
\end{equation}
The first solution is the trivial perturbative solution, while the
second, nontrivial non-perturbative solution determines
$\sqrt{\Lambda}$ in terms of the bare gravitational coupling
constant $G_0$ and the cutoff $K_c$. The nontrivial solution
corresponds to the superfluid condensate state which is the true
vacuum state of the system, while the trivial solution corresponds
to the normal (false) vacuum state i.e. not to the true vacuum
state \footnote{A de Sitter space solution to the non-perturbative
gap equation has been obtained for a large $N$ expansion by
Inagaki et al.~\cite{Inagaki}.}.

The gap equation (\ref{gap}) asymptotically has an exponential
dependence on the physical density of states.  We have that in an
FRW background filled with fermions the energy gap will take on
the following form, $\Delta=\sqrt{\Lambda}$, separating the two
phases. Physically this means that the gap corresponds to the
binding energy necessary to form the condensate. The difference
between the original vacuum energy and the final vacuum energy is
the rest mass of the condensate.

\section{Conclusions}

We have shown that the vacuum energy in the early universe can
become unstable as the attractive fermion-gravitational force and
a screened attractive force between fermions produces condensates
through a phase transition at a critical temperature $T < T_c$.

The non-perturbative phase transition generates a gap in the
ground state or vacuum energy and we make the identification
$\Delta\sim \sqrt{\Lambda}$. An initial phase of de Sitter
inflation caused by a large vacuum energy (cosmological constant)
is followed by an exponential suppression of the cosmological
constant as the universe expands. We described this scenario by
analogy with the BCS mechanism associated with the formation of
Cooper pairs of fermions by means of the exchange of phonons,
replacing the phonon exchange by graviton exchange.

The non-perturbative mechanism can explain how an initially large
vacuum energy (cosmological constant) can be suppressed by a phase
transition to a superfluid state of the early universe as the
universe expands, leading to a ``graceful'' exit for inflation.

The non-relativistic model we have used to describe the scenario
can be extended to a relativistic QFT model of the formation of a
vacuum energy gap for fermion condensates in a de Sitter spacetime
background.

This scenario explains why a naive {\it perturbative} calculation
of the vacuum energy leads to a nonsensical answer. In contrast to
the ground state or vacuum of QED or the standard model, the
vacuum associated with gravity is unstable and the instability can
only be described by non-perturbative physics.

\section*{Acknowledgments}

This work was supported by the Natural Sciences and Engineering
Research Council of Canada. I wish to thank my collaborator
Stephon Alexander for stimulating and helpful discussions.

\end{document}